\documentstyle[aps,multicol,epsf,amstex,amssymb]{revtex}

\draft

\def\half{{{1\over2}}}
\def\squote{}
\def\quote#1#2#3#4{\squote {#1,\ {\sl#2}\ {\bf#3}, #4}.\par}
\def\qquote#1#2#3#4{\squote {#1,\ {\sl#2}\ {\bf#3}, #4};}

\def\prl{{\sl Phys. Rev. Lett.}\ }

\def\pr {{\sl Phys. Rev.}\ }

\def\w{\omega}
\def\e{\epsilon}
\def\deriv{\partial}

\begin{document}
\title{Two-species percolation and Scaling theory
  of the metal-insulator transition in two dimensions}
\bigskip
\author{\large  Yigal Meir}
\address{Department of Physics, Ben-Gurion University, Beer Sheva 84105, ISRAEL}
\maketitle
\begin{abstract}
Recently, a simple non-interacting-electron model,  combining
local quantum tunneling via quantum point contacts
and global classical percolation,  has been introduced in order to
describe the observed ``metal-insulator transition'' in two
dimensions \cite{mywork}. Here, based upon that model,
a two-species-percolation scaling theory
 is introduced and compared to the experimental
data. The two species in this model are, on one hand, the
``metallic'' point contacts, whose critical energy lies below the
Fermi energy, and on the other hand, the insulating quantum point
contacts.  It is shown that many features of the experiments,
such as the exponential dependence of the resistance on
temperature on the metallic side, the linear dependence of the
exponent on density, the $e^2/h$ scale of the critical resistance,
the quenching of the metallic phase by a parallel magnetic field
and the non-monotonic dependence of the critical density on a
perpendicular
 magnetic field,  can be naturally explained by the model.
 Moreover, details such as the nonmonotonic dependence of
 the resistance on temperature or the inflection point of the
resistance vs. parallel magnetic are also a natural consequence of
the theory. The calculated
parallel field dependence of the critical density agrees excellently
with experiments, and is used to deduce an experimental value of the
confining energy in the vertical direction.
It is also shown that the resistance on the
``metallic'' side can decrease with decreasing temperature by an
arbitrary factor in the degenerate regime ($T\lesssim E_F$).

\end{abstract}
\begin{multicols}{2}
\section{Background and introduction of the model}
The surprising experimental observation of a metal-insulator
transition in two dimensions
\cite{kravchenko,coleridge,hanein1,hanein3,yaish},
in contradiction
with the predictions of single-parameter scaling theory for
noninteracting electrons \cite{abrahams}, has been a subject of
extensive investigation in recent years. Theories ranging from
attributing the effect to  scattering by impurities
\cite{altshuler} to those suggesting ``a new form of matter''
\cite{pichard} have been proposed.
  Some theories, based on the treatment
of disorder and electron-electron interactions  by Finkelstein
\cite{finkelstein}, have been put forward \cite{interactions},
while other approaches considered spin-orbit scattering
\cite{pudalov_model} or  percolation of electron-hole liquid
\cite{he}. Altshuler et al. \cite{altshuler} gave several
arguments why this transition is not due to a non-Fermi liquid
behavior, including the fact that the exponential increase of the
conductance with temperature persists to high densities where the
conductance is almost two orders of magnitude larger than the
critical conductance, and the fact that the Hall resistance is
rather insensitive to temperature, and does not display any
critical behavior. Some experimental results supporting the
conclusion that the transition is not driven by interactions that
were mentioned in \cite{mywork} included the fact that such a
transition was observed also in high-density electron gas upon the
introduction of artificial disorder \cite{ennslin} and the fact
that increasing the density in a parallel electron gas increases
the conductance \cite{yaish}, even though the  interactions are
screened by the parallel gas. More recently,  the compressibility on
the metallic side of the transition was measured \cite{amir} and
was shown to
 be accurately described by Hartree-Fock approximation, again
 indicating
a normal Fermi-liquid behavior. Several other recent 
experiments \cite{pepper,senz} have demonstrated weak-localization
behavior on the metallic side with very little effect of
electron-electron interactions.

Recently \cite{mywork} I proposed a simple  non-interacting
electron model,  combining local quantum tunneling and global
classical percolation,  to explain several features of the
experimental observations. At low electron or hole densities
  the potential fluctuations due to the
 disorder cannot be screened and they
define density puddles  (density separation into puddles
 in gated GaAs was indeed observed experimentally by Eytan et al.
 \cite{israel},  using near-field spectroscopy).
 These puddles are connected via saddle points,
 or quantum points contacts (QPCs).
It is now established that
 even at low temperatures and for open puddles (or quantum dots),
 the dephasing time may be shorter than the escape time from
 the puddle \cite{marcus}. Thus it is assumed that  between
  tunneling events through the QPCs dephasing takes place,
and  the conductance
 of the system will be determined by adding classically these quantum
 resistors.  (A related model was introduced by Shimshoni et al. \cite{efrat}
 to describe successfully transport in the quantum Hall  regime.)
 Each saddle point is characterized
 by its critical energy $\e_c$, such that the
 transmission through it is given by $T(\e)=\Theta(\e-\e_c)$.
 (I assume that the
 energy scale over which the transmission changes from zero to
unity is smaller than the other relevant energy scales,  to avoid
additional parameters).
 Then
 the conductance through a QPC is given by the Landauer formula,
\begin{eqnarray}
G(\mu,T) &=& {2e^2\over h} \int d\e
 \left(-{{\deriv f_{FD}(\e)}\over{\deriv \e}}\right)
 T(\e)\nonumber\\
  &=&  {2e^2\over h} {1\over{1+\exp[(\e_c-\mu)/kT]}} ,
\label{QPC}
\end{eqnarray}
where $\mu$ is the chemical potential,  and $f_{FD}$ is the
Fermi-Dirac distribution function.

The system is now composed of classical resistors,  where the
resistance of each one
 of them is given by (\ref{QPC}),  with random QPC energies.

\section{Two-species scaling theory}
In \cite{mywork} I presented numerical calculations to be compared
to the experimental data. Here I present a different approach,
based on the scaling theory of a two-species percolation network.
At low temperatures the resistors can be divided into two groups,
the conducting ones ($\e_c<\mu$), whose conductance is about
$2e^2/h$, and the insulating ones ($\e_c>\mu$), whose conductance
is nearly zero. Thus the distribution of the conductances will be
a two-peak distribution, where the weight of each peak will be
 determined mainly by density (or chemical potential) and the
 position of the conducting peaks will be determined mainly by
  temperature. Since most properties of such a percolating
network are insensitive to details of this distribution,
 I replace it by a two-delta function distribution, namely
 replace the network by a network comprising of two types
 of conductors: \\
 An effective conductor $\sigma_m$ describing the metallic QPCs, and
 whose conductance is given by (\ref{QPC}), with an
 appropriately averaged $\e_c$:
 \begin{equation}
 \sigma_m = {2e^2\over h} {1\over{1+\exp[-A/kT]}}
\label{sigmac}
\end{equation}
($A$, which depends on the potential fluctuations distribution,
  is taken as unity in the following, i.e. it defines the temperature scale),
and  an effective conductor $\sigma_i$ describing the contribution
of the insulating phase. The conductance of the insulating
QPC is dominated by activation\cite{vrh},
$ \sigma_i = \sigma_a \exp[-A_1/T]$.
(Indeed, experimental invesigations reported that
``two different contributions to the conductivity (or two conducting
systems) may exist, one with a metallic temperature behavior and another
one with a standard, insulating, weak-localization behavior'' \cite{senz}.)

The scaling form of the two-dimensional
conductance of such a two-phase
mixture, $\sigma$, near the percolation threshold is well known
\cite{scaling,stauffer},

\begin{equation}
\sigma = \sqrt{\sigma_m\sigma_i}\ f\left[(n-n_c)^t
\sqrt{\sigma_m/\sigma_i} \right]
\label{scaling}
\end{equation}
with $t\simeq1.3$, the conductance critical exponent for two-dimensional
percolation, and
\begin{equation}
f(x) \propto
\begin{cases}
x & x\rightarrow \infty\\
1/x & x\rightarrow -\infty
\end{cases}\ \  \ ,
\label{f}
\end{equation}
 so that in the case $\sigma_i\rightarrow0$
  (a regular random resistor network),
 $\sigma(n)\sim \sigma_m(n-n_c)^t$, while in the case
 $\sigma_m\rightarrow\infty$ (a mixture of an insulator and
  a superconductor),
 $\sigma(n)\sim \sigma_i(n_c-n)^{-t}$. (In the above I used
  the notation $x^t\equiv sign(x) |x|^t$.) The exact form of
  $f(x)$ is not very important, and in the following I have
  chosen $f(x)=\log(B+\exp[x])/\log(B+\exp[-x])$, with $B=2$.

\section{``Zero'' temperature}

 At low enough temperatures, such that $T\ll A,A_1$
the conductors have either zero
conductance or a conductance
 equal to $2e^2/h$. If the dephasing time is still finite at these
temperatures,  one has a random-resistor network,
 which exhibits a second-order
  percolation transition\cite{stauffer}.
  In Fig.~1 I fit the lowest-temperature experimental data of
 \cite{hanein1,hanein3,yaish} to the
 expected critical  dependence.
 Clearly, the agreement with the classical percolation  prediction
 is excellent. Such as agreement with the classical percolation critical
behavior may serve as an additional experimental indication that
the dephasing is still finite at the lowest available experimental
temperatures.

\vskip  -0.5 truecm
\begin{center}
\leavevmode \epsfxsize=3.5in
\epsfbox{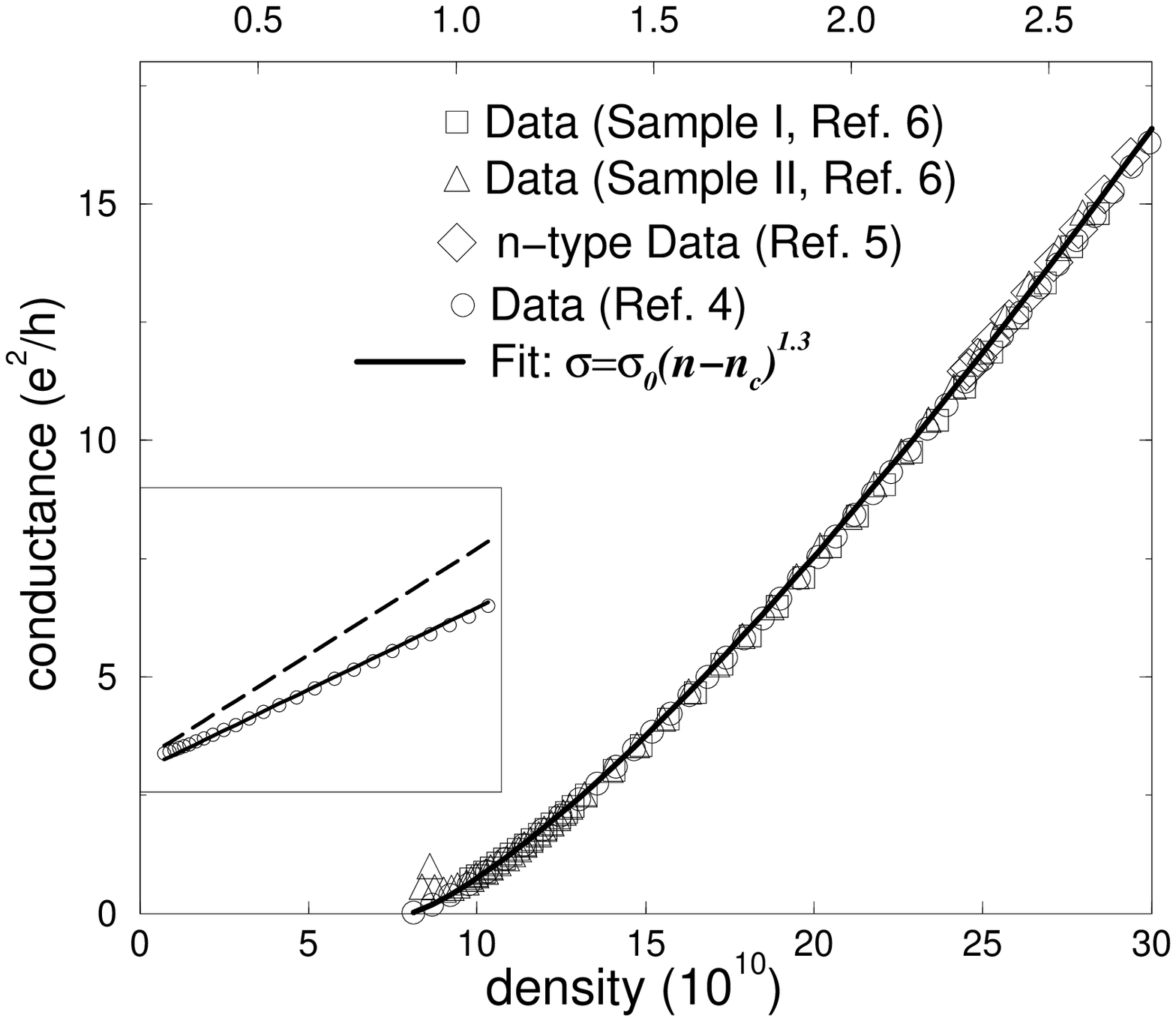}
\end{center}
\begin{small}
\vskip -0.5 truecm
Fig. 1. Comparison of the lowest temperature data of \cite{yaish}
 (two sets of data, triangles and squares,  330mK,
 density given by the lower axis)
and of \cite{hanein1} (circles,  57mK, density given by the upper
axis), and of the n-type data  \cite{hanein3} (diamonds)
  to the prediction of
 percolation theory (solid line). Inset: Logarithmic derivative
  of the data
\cite{yaish} which gives a line whose slope is inverse of the
critical exponent. The percolation prediction $(t\simeq1.3)$ is given
by the solid line. For comparison a $t=1$ slope is also shown
(broken line).
\end{small}
\vskip 0.5 truecm

In the inset I plot the experimental data\cite{yaish} for
$1/(d\log\sigma/d\log n)$, which, if indeed $\sigma\sim(n-n_c)^t$,
is given by $(n-n_c)/t$. The data indeed fits on a straight line,
with a slope given by $1/t\simeq1/1.3$. For comparison a straight
line with a slope of unity is also depicted, in order to
demonstrate that a critical exponent of unity cannot fit the data.

\section{Temperature dependence of the resistance}
As temperature increases,  the Fermi-Dirac distribution is
broadened. Consequently the conductance of the  of the transparent
quantum point contacts ($\e_c<\mu$) decreases  exponentially (towards half its
value),
while that of the insulating ones increases. Thus we expect to see
rather dramatic effects as a function of temperature.
 This is
indeed depicted in Fig.~2. In (a) we plot the prediction of the
model and in (b) the experimental data \cite{hanein1}. As
temperature is lowered, systems with slightly different resistance
at high temperatures will diverge exponentially with decreasing
temperatures. The resistance of systems on the metallic side
($n>n_c$) will saturate at zero temperature, while that of
insulating   samples will diverge, in agreement with the general
shape of the experimental curves.
 Note that there is an upward
turn even on the metallic side of the transition. In fact,  close
to the transition,  on
 the metallic side,  as temperature decreases, the conductance of the
 insulating part decreases significantly, and its
 contribution to the total conductance is dramatically reduced.
  Since the critical percolation cluster is very ramified (in
 fact of fractal dimension),  the contribution of the insulating part
 of the system dominates at high temperatures,
   and the increase of its resistance with decreasing temperature
leads to an increase of the resistance as temperature is lowered,
even on the metallic side. At low enough temperatures, however,
when the resistance of the insulating part of the network becomes
high enough,  its contribution to the total conductance becomes
negligible. Then the total resistance is dominated by the
percolating conducting network, and thus the overall resistance
will decrease with decreasing temperature. This leads to a
nonmonotonic temperature dependence on the metallic side of the
critical point, which can be clearly observed both in the model
(Fig. 2c) and in the data (Fig. 2d).

The fact that only deep in the metallic regime, the overall
resistance increases with increasing temperature, suggests that the
density at which the resistance is approximately temperature
independent is not the true critical point,  but rather deeper on
the metallic side. This is clearly seen in Fig.~3b,  where one can
see a  point where all the low-temperature curves nearly cross,
well inside the metallic regime. The above discussion suggests
that one should be cautious in associating the critical point with
the experimentally observed ``temperature-independent'' point
(Fig.~3a),
 as done routinely in the experiment interpretations.

\vskip  -0.5 truecm
\begin{center}
\leavevmode \epsfxsize=3.5in
\epsfbox{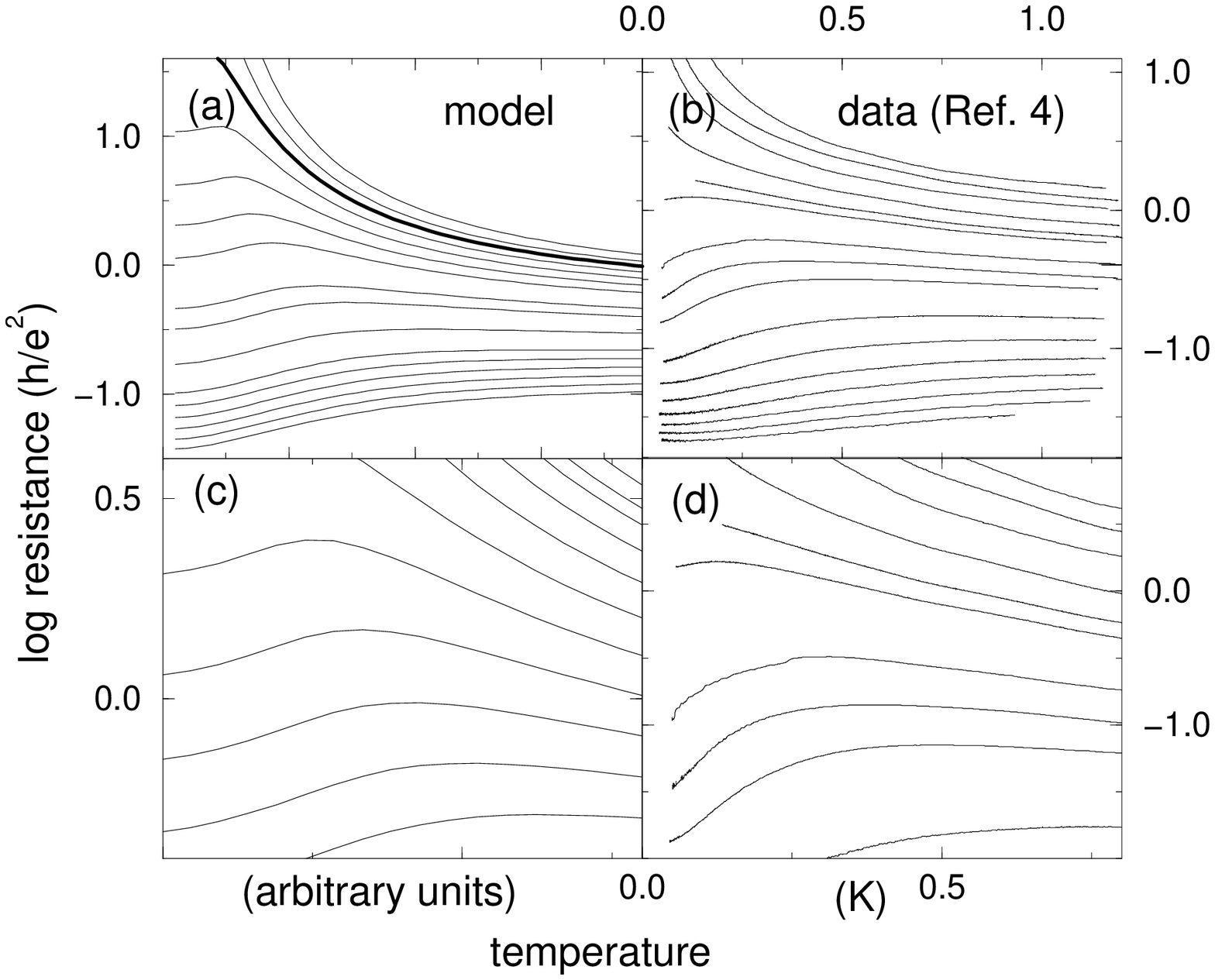} \vskip -0.5 truecm
\end{center}
\begin{small}
Fig. 2. Temperature dependence of the resistance for systems of
different densities, as obtained by the model (a) and compared to
the experimental data of Ref.\cite{hanein1} (b). The critical line
is denoted by the bold curve. All curves below the critical line
saturate at zero temperature, while above it the resistance
diverges. For systems close to the transition on the metallic
side, the resistance is a nonmonotonic function of temperature, as
seen both in the model (c) and in the data (d).
\end{small}

\begin{center}
\leavevmode \epsfxsize=3.5in
\epsfbox{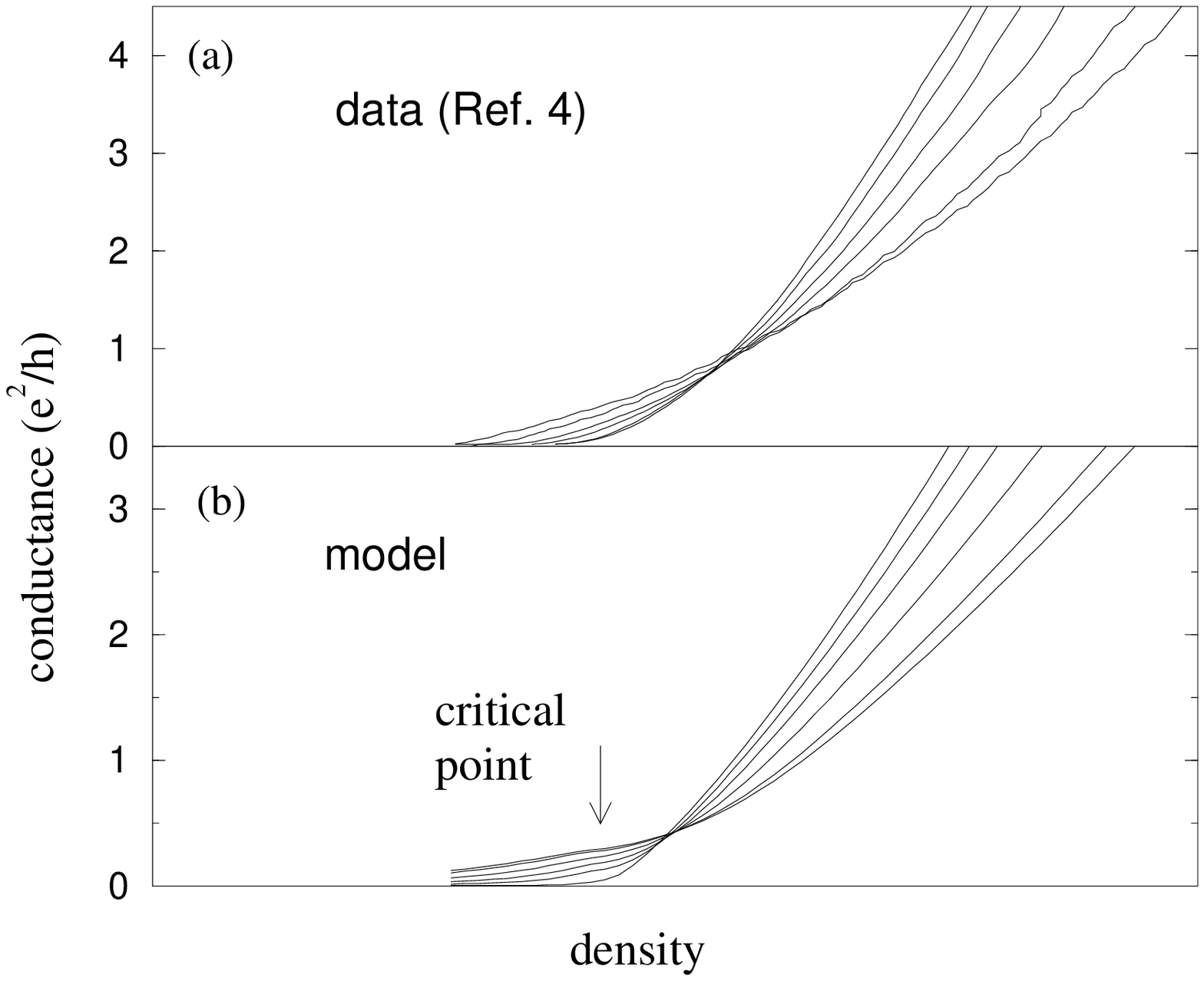}
\end{center}
\vskip -0.5 truecm
\begin{small}
Fig. 3. Comparison of the density dependence of the conductance between
the data of \cite{hanein1} (a) and the model (b), for several
temperatures. The density at which the theoretical curves seem to
cross each other, is well above the true critical point.
\end{small}
\vskip 0.5 truecm

Lastly, The resistance at the metallic regime is given by some
geometrical factor times the inverse of $\sigma_m$
(Eq.~\ref{sigmac}), which naturally gives the observed exponential
temperature dependence observed experimentally. The
high-temperature resistance of the critical density network is
naturally around $h/e^2$,  the only resistance scale in this
model.

\section{Parallel magnetic fields}
The effect of a parallel magnetic field on the overall conductance
is determined by the way it affects the individual points
contacts. The effect of a
 parallel  field on transport through a single QPC has been studied
 in detail
 \cite{qpc_parallel}. These experimental and theoretical
  studies demonstrated that the
 threshold density  where the QPC opens up {\sl
 increases} parabolically with the in-plane magnetic field. This
  effect was  attributed to  coupling of
 the in-plane motion to the strong confinement in the vertical
 direction,  leading to an increase in the confining enrgy. Writing,
 for simplicity, the three dimensional Hamiltonian that describes
  free motion in two-dimensions and a harmonic confining potential
  in the third ($z$) direction, with a magnetic field pointing in
  the $x$-direction
 \begin{equation}
 {\cal H} = {{p_x^2}\over{2m}} + {{(p_y+eBz)^2}\over{2m}} +
 {{p_z^2}\over{2m}} + \half m \omega_0^2 z^2\ \ \ ,
 \label{H3d}
 \end{equation}
it is straightforward to see that the bottom of the 2d band shifts
from $\hbar \omega_0/2$ to $\hbar \sqrt{\omega_0^2+\omega_c^2}/2$,
with $\omega_c\equiv eB/mc$, leading to a corresponding decrease
in the kinetic energy of all electrons. Thus the effective
critical Fermi energy, or density, becomes larger.
In other words, for a given density or chemical potential, 
if the system at zero field is on the metallic side, i.e.
  if the Fermi momentum is above the critical
momentum (or kinetic energy) allowing percolation through the
system, a parallel field will lower that energy towards the
critical energy, eventually crossing the critical point and
leading to an insulating behavior. Fig.~4 depicts the
experimental data of \cite{yoon}, and the corresponding predictions
of the model. As expected, as magnetic field increases, the system
gradually crosses over from a metallic to an insulating behavior.

The above discussion allows a quantitative prediction of the
effective critical energy in a parallel field ,
\begin{equation}
\label{Hc}
  \e_c(H) = \e_c(H=0) +
 \hbar \left( \sqrt{\omega_0^2+\omega_c^2}-\omega_0 \right)/2  .
\end{equation}

At zero temperature, when the resistance on the insulating side is
infinite, we expect the resistance on the metallic side
$\mu>\e_c(H=0)$ to diverge with increasing field,
 \begin{equation}
R(H)\sim
(\mu-\e_c(H))^{-t} .
\end{equation}

\begin{figure}
\begin{center}
\leavevmode
\epsfxsize=8truecm
\epsfbox[53 13 639 780]{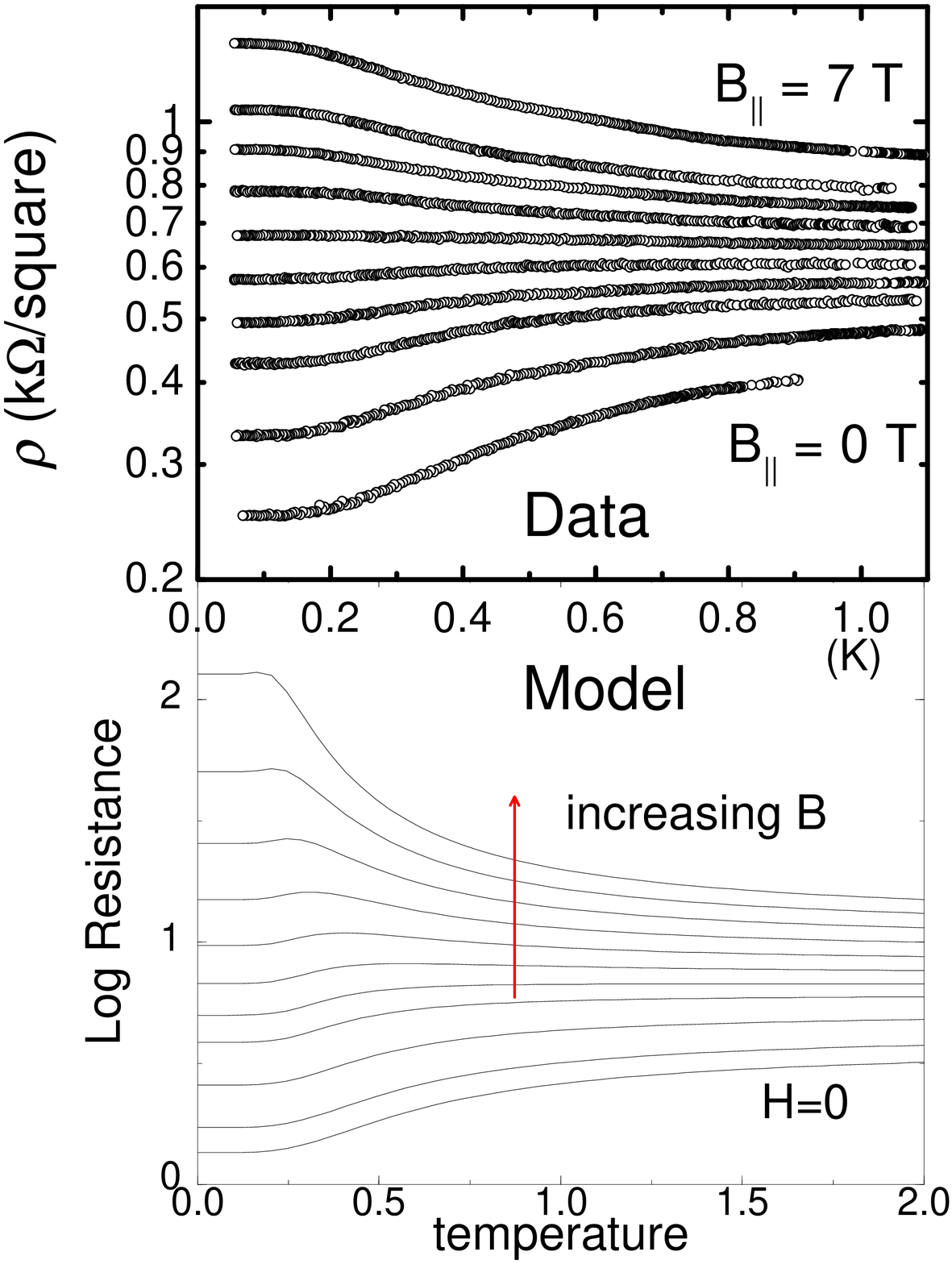}
\end{center}
\begin{small}
Fig.~4. Comparison between the experimental data
 \cite{yoon}
and the predictions of the model,  demonstrating that
 a parallel magnetic field
causes the metallic system to cross over gradually to the insulating
regime.
\end{small}
\label{ResTH}
\end{figure}

 For a finite temperature
this divergence is cut off by the finite resistance of the
insulator, and the magnetic field dependence changes as one
crosses into the insulating side. This behavior is clearly seen in
the experimental data of \cite{yoon}
(very similar data was also reported by Mertes et al. \cite{mertes}).
Fig.~5
 depicts the experimental data \cite{yoon}
for the magnetic field dependence of the resistance, compared to
what is expected from the model. The difference in behavior
between the metallic and the insulating regimes is clear. On the
metallic side we see that the resistance increases rapidly as the
magnetic field brings the critical point closer to the chemical
potential and then an abrupt change of behavior as the system
enters the insulating regime. If the system was on the insulating
side to begin with, then the magnetic field dependence of the
resistance is similar to what is usually seen in systems where
transport is via variable-range hopping \cite{zvi}. In these
system the positive magnetoresistance is due to spin polarization
\cite{kamimura}.
 The magnetic field dependence on the insulating
side  depicted in Fig.~\ref{yoonHn} is assumed here to be due to that
process. In this regime, however, the magnetic field at which
there is a marked change in behavior is spin-related and
should depend only weakly on density. A recent
experimental investigation of the magnetoresistance in the
insulating side \cite{shlimak} indeed supports this mechanism.

\begin{center}
\leavevmode
\epsfxsize=3.5in
\epsfbox[53 203 639 950]{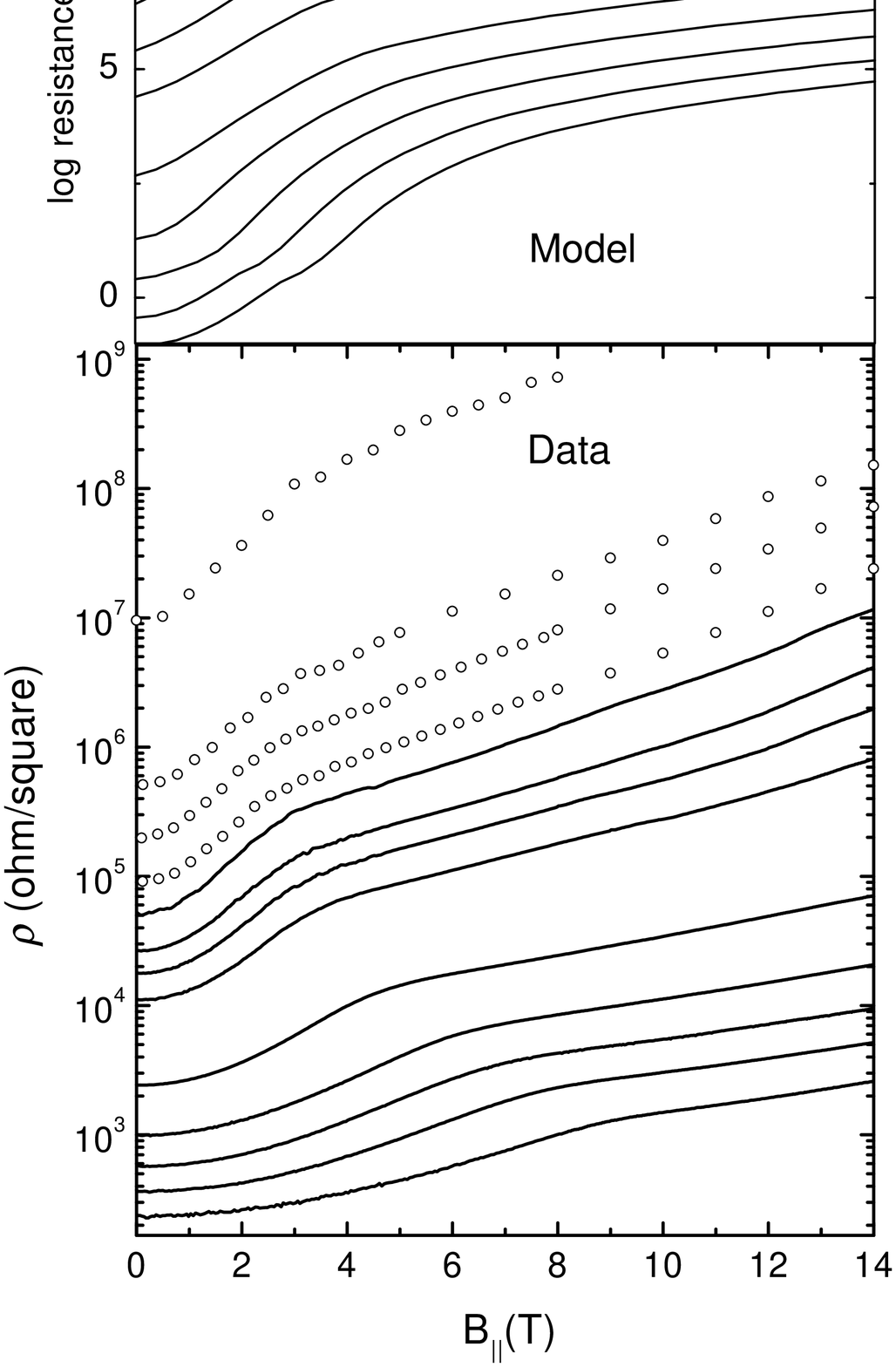}
\end{center}
\vskip 3 truecm
\begin{small}
Fig.~5.
Comparison of the experimentally measured resistance, as a function
of parallel magnetic field \cite{yoon} to the model predictions. On the
metallic side, the magnetic field shift the critical point towards the
chemical potential, leading to a divergence in the resistance, which is cut off
by finite temperature.
\label{yoonHn}
\end{small}
\vskip 0.5 truecm

As was mentioned above, the critical point in the density -- magnetic field
plane shifts towards higher densities with increasing magnetic field.
Yoon et al. \cite{yoon}
have also measured the dependence of the critical field
on density, which can be deduced from the inversion of Eq.(\ref{Hc}),

\begin{equation}
\label{hcfield}
 H_c =  m^*c/e \sqrt{\Delta(H)^2+2 \hbar\omega_0 \Delta(H)} , 
\end{equation}
where $\Delta(H)\equiv\e_c(H)-\e_c(H=0).$ The comparison of the
prediction of this simple equation to the data is depicted in
Fig.~6 for two samples cut from the same wafer. The fitting
parameters are the zero-field critical point, that can be read
directly from the data, the gate capacitance - the rate at which
the Fermi energy changes with density, and the confining energy in
the perpendicular direction, $\hbar \w_0$. It is encouraging to
note that while the critical energy, which is determined by the
disorder realization, and the gate capacitance, which is
determined by the geometry, are different for the two samples,
both sets of data can be fitted by the same value of
 the perpendicular confining energy, which ought to be the same
for the two samples, and turns out to be
 $\hbar \w_0\simeq 0.8 meV$, leading to an
extension of the wavefunction in the perpendicular direction of
the order of 11 nm, similar to the value used by the authors of
Ref.\cite{amir} to fit their experimental data.

\vskip -0.5 truecm
\begin{center}
\leavevmode \epsfxsize=3.5in
\epsfbox{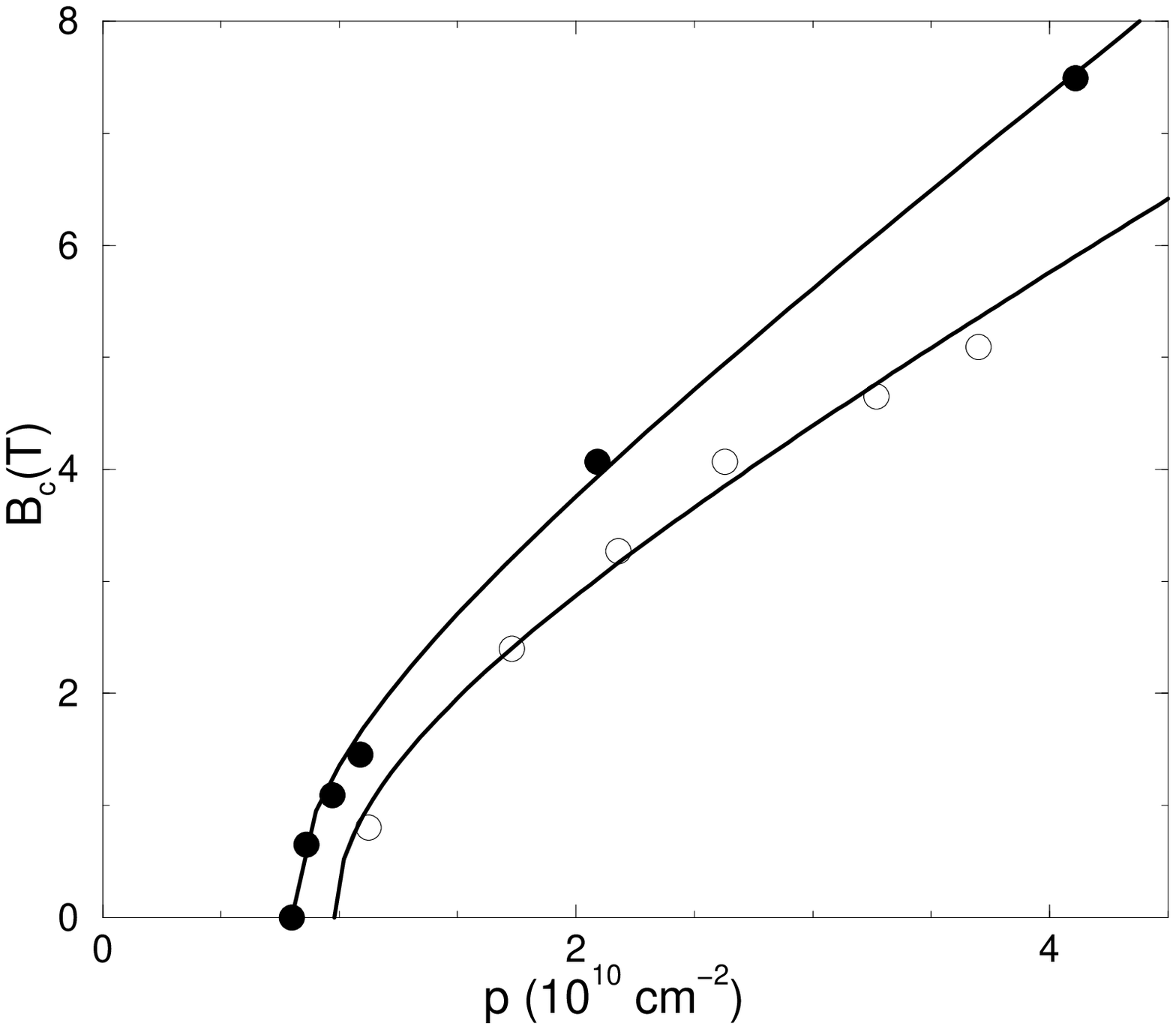}
\end{center}
\vskip -0.5 truecm
\begin{small}
Fig. 6. Comparison of the measured density dependence of the
critical magnetic field \cite{yoon} (circles) to the prediction of
the theory (Eq.\ref{hcfield}, solid lines). The two data sets are
two different samples cut from the same wafer, and can be fitted
using the same value of the perpendicular confining energy.
\label{yoonHc}
\end{small}
\vskip 0.5 truecm

Interestingly, it seems that the effects of parallel fields can be
understood without employing the electron spin. As a parallel
field will also reduce the conductance of some of the  point
contacts from $2e^2/h$ to $e^2/h$, the Zeeman effect will also
increase the system's resistance.
 It should be
noted that while the coupling of the in-plane motion to the
confining potential in the perpendicular direction was also
considered by Das Sarma and Hwang \cite{dassarma},
 the magnetoresistance predicted here,  in contrast with
 Ref. \cite{dassarma},  should not exhibit any anisotropy.
  The reason
  is that the direction of transport through the quantum point contacts
 is expected to be random,  with no preferred direction.

\section{perpendicular magnetic fields}
While the longitudinal resistance depends exponentially on
temperature, the weak field Hall resistance is practically
independent of temperature \cite{pudalovHall}. Such an observation
might be hard to account for in theories that argue for new
non-Fermi liquid-like behavior, but it is trivial in the present
model - the critical exponent for the Hall coefficient in a two
dimensional percolation problem is exactly zero \cite{shklovski},
and thus the Hall coefficient should display no critical behavior
at the critical point. This prediction was indeed confirmed in
classical percolation experiments \cite{palevski}, and
 is very similar to that observed by Pudalov et al. \cite{pudalovHall}.

\vskip -0.5 truecm
\begin{center}
\leavevmode \epsfxsize=3.5in
\label{interlevel}
\epsfbox{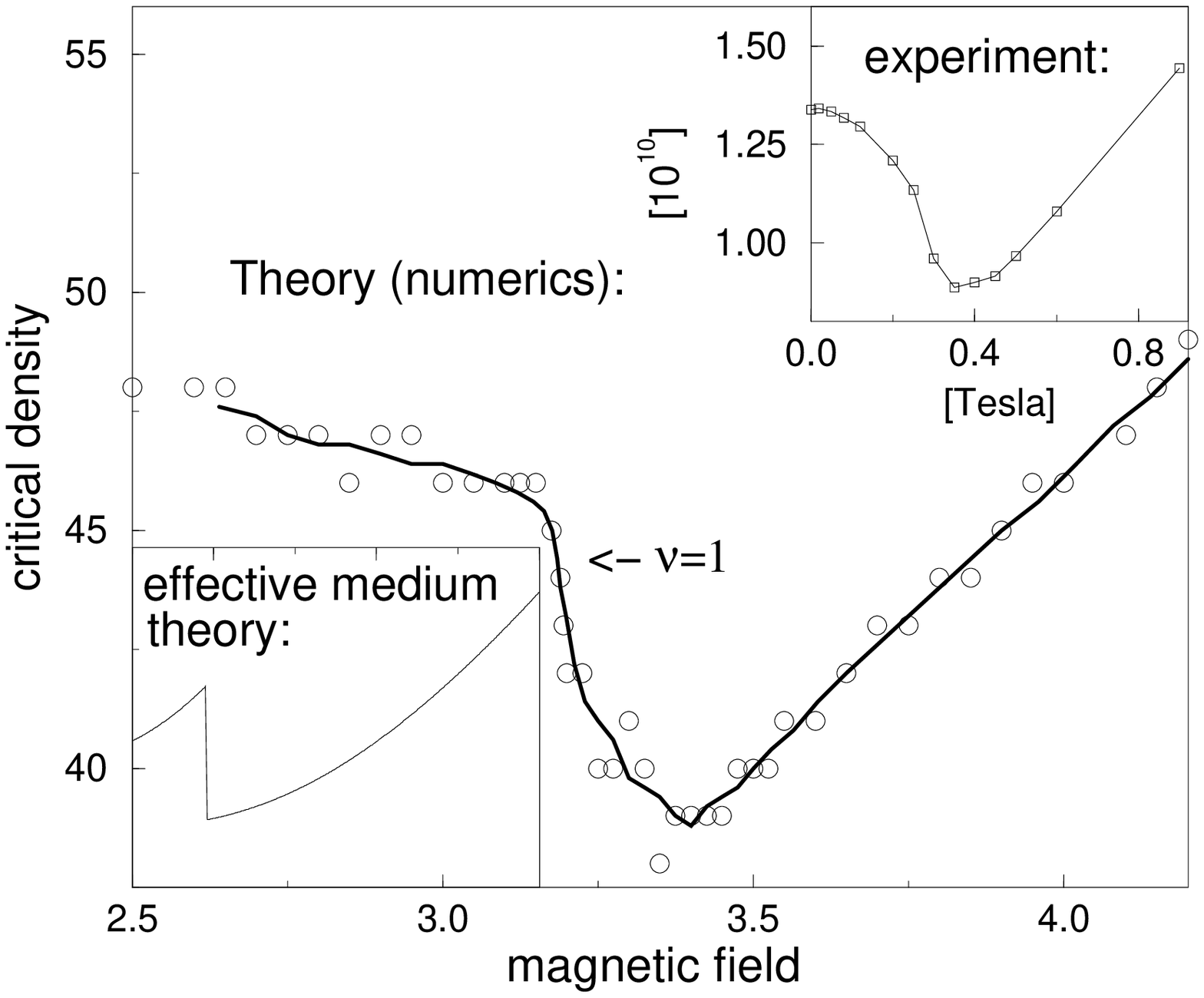}
\end{center}
\vskip -0.5 truecm
\begin{small}
Fig. 7. The critical density - the number of electrons in the puddle,  so that
the topmost energy will allow transport through the point contact - as a
 function of magnetic field,  in the presence of a finite disorder. The
continuous curve is an averaged fit through the (necessarily
integer) data points. Top-right inset: the corresponding
experimental data \cite{magfield}. Bottom-left inset: results of
effective medium theory \cite{myEMT}.
\end{small}
\vskip 0.5 truecm

The situation in larger  perpendicular magnetic fields is more
interesting,  as quantum Hall (QH) states are formed. Transport
through a single QPC in perpendicular field and the crossover
between the zero field limit and the QH limit have been studied in
detail \cite{vanwees}. As expected, one finds that the critical
energy oscillates with magnetic field due to the depopulation of
Landau levels. In the present case, the oscillations are smoothed
out by the disorder and by the averaging over many QPCs. Thus only
the strongest oscillation,  near  $\nu=1$,  may survive, leading
to a single dip in the critical density vs. magnetic field plot,
as was observed experimentally. In order to allow for the
averaging procedure, one has to take the full conductance
distribution into account, which is beyond the two-species scaling
 theory. For completeness I report here results of numerical
 calculations \cite{mywork} and effective medium theory
 \cite{EMT,myEMT}.
In the  numerical calculation I studied the energy levels of one
puddle of electrons,  which we modeled by a circular disk,  in the
presence of disorder \cite{disorder}. In Fig.~7 I plot the
``critical density'' -- the number of electrons that need to
occupy the puddle,  so that the energy of the highest-energy
electron will be enough to transverse the QPC \cite{fertig},
equivalent in the bulk system to the critical density -- as a
function of magnetic field. Indeed  a dip near $\nu=1$ is clearly
seen,
with all other oscillations smoothed out by the disorder. This
curve has a strong resemblance to the experimental data
\cite{magfield} (top-right inset). The results of the effective
medium theory are depicted in the bottom-left inset, again
demonstrating a dip near $\nu=1$.
 In addition, it is  expected that as the magnetic field is
lowered below the $\nu=1$ minimum, more than one channel will
transverse some QPCs, leading to an increase in the critical
conductance,  as indeed reported experimentally.

\section{Reduction factors larger than 2}
In the degenerate electron limit ($T\ll E_F$), the biggest
reduction factor in the resistance on the metallic side, with
decreasing temperature is a factor of $2$. On the other hand,
reduction factors close to an order of magnitude and even larger
\cite{newpudalov} have been observed experimentally, especially in
Silicon based samples . Here I show that when temperature becomes
of the same order of magnitude of $E_F$, the reduction factor in
the model can assume arbitrarily large factors \cite{paulmu}.

The electron density is given by
\begin{equation}
\label{nmu}
  n = \int_0^\infty d\e {{\rho_0}\over{1+\exp[(\e-\mu)/T}]} =
  \rho_0\ T \log \left[1+\exp(\mu/T)\right]
\end{equation}
where $\rho_0$, assumed constant, is the electronic density of
states (per energy and per volume), and $\mu$, the chemical
potential, is measured relative to the bottom of the band.
Inverting the above equation, the chemical potential for a given
density, is
\begin{equation}
\label{mun}
  \mu = T \log \left[\exp(n/\rho_0 T)-1\right] \equiv
   T \log \left[\exp(E_F/T)-1\right]
\end{equation}
where the Fermi energy is defined as the $T\rightarrow0$ limit of
the chemical potential. These textbook expressions demonstrate
that while the Fermi energy varies linearly with the density, the
chemical potential may be more sensitive to density variations in
the nondegenerate limit $T\sim E_F$. Moreover, the chemical potential
is now temperature dependent, and decreases with increasing temperature
(see inset in Fig.~8).
Substituting the above expression in the expression
for the conductance through a  single point contact (\ref{QPC})
demonstrates that the conductance can decrease arbitrarily with increasing
temperature (or,  equivalently,  that the resistance can decrease by
an arbitrary factor with decreasing temperature).
  In Fig.~8 the
temperature dependence of the conductance is plotted for two
values of the Fermi energy,  $E_F$. The curve for $E_F\gg T$ shows
the expected behavior for the degenerate electron gas - the
conductance decreases and
 saturates at a value smaller from the zero temperature conductance by
 a factor of 2. On the other hand, in the nondegenerate regime,
  $E_F\sim T$, the conductance decreases by a much
  larger factor.

\vskip -0.5 truecm
\begin{center}
\leavevmode \epsfxsize=3.5in
\label{muTfig}
\epsfbox{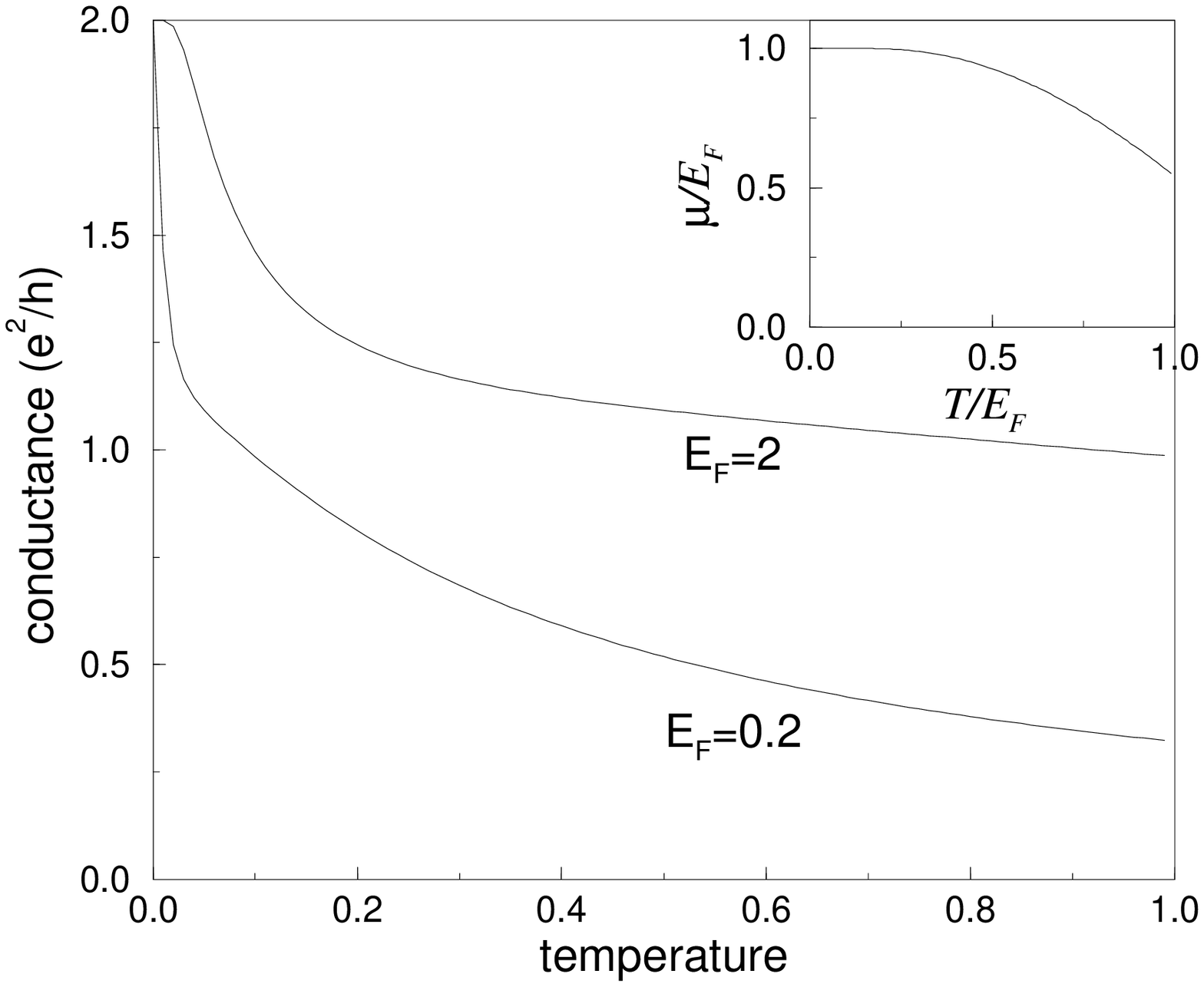}
\end{center}
\vskip -0.5 truecm
\begin{small}
Fig. 8. The temperature dependence of the conductance in the degenerate
($kT\ll E_F$) and in the degenerate regime ($kT\simeq E_F$). While in
the nondegenerate regime, the conductance can decrease by a factor of
two, it can decrease arbitrarily in the nondegenerate regime, due to
the temperature dependence of the chemical potential (inset).
\end{small}
\vskip 0.5 truecm

\section{Conclusions}

All the above results and discussion demonstrated that many of the
 experimental observations can be explained in the context of the
 simple semi-classical, noninteracting model introduced
here. This is not to say that interactions and other effects are
irrelevant. For example, the formation of the electron puddles may
be dominated by interaction effects (see, e.g., Ref.\cite{he}),
and dephasing is certainly dominated, at these low temperatures,
by electron-electron interactions. Other effects, including the
energy dependence of the transmission coefficient and the
possibility of more than one channel through the QPCs, the role of
interband-scattering \cite{yaish} and temperature-dependent
impurities \cite{altshuler} may also be important to understand
quantitative aspects of
 the data. Nevertheless,  the fact that several important aspects
 of the experimental data can be explained in the context of
 a simple model is quite  encouraging.

Some predictions made in \cite{mywork}, where the model was
presented, were already confirmed. The mechanism for the quenching
of the metallic phase by a parallel magnetic field was suggested
there, and, as discussed above, agrees very well with recent
experiments. Moreover, it was suggested that local measurements
will be able to explore the percolative nature of the insulating
phase. Indeed, Ilani et al. \cite{amir} have used local probes to
measure the change of the local chemical potential with density.
While on the metallic side the signals from all probes were
identical, and were accurately described by Hartree-Fock theory,
these probes gave different signals on the insulating side, which
the authors interpreted as a signature of a percolative phase. (An
indirect experimental verification of the percolation process in
the QH regime was already reported in  \cite{qh}).
 As the metallic puddles can be thought of as quantum
dots,  one can use the abundant information about  such structures
\cite{qdots}, to gain additional understanding of the
characteristics of the puddles and the phase separation. Such
local probes \cite{ashoori} can give a "smoking gun"  verification
of the picture presented
 here by looking for the  periodic oscillations of the local chemical
 potential on the insulating side,
due to depopulation of the Landau levels,  as was observed in
 quantum dots  \cite{paul}.

 To conclude, a semi-classical model, combining local quantum
 transport and global classical percolation was employed to
 explain the observed metal-insulator transition. The model
 attributes the transition to the finite dephasing length in these
 temperatures. As temperature is lowered even further, and the
 dephasing length becomes larger than the puddle size, quantum
 localization effects should kick in. Such weak-localization
 corrections in the metallic side were indeed observed
 experimentally, confirming the expectation that, if the dephasing
 length indeed diverges at zero temperature, these systems will
 eventually  becomes Anderson insulators.

 I thank many of my colleagues for fruitful discussions:
 A.~Auerbach, Y.~Gefen, Y.~Hanein, P. McEuen, D.~Shahar, E.~Shimshoni,
 U.~Sivan,
 A.~Stern, N.~S. Wingreen,  A.~Yacoby and  Y. Yaish. In particular, I would like to
thank Y. Hanein \& D. Shahar and U. Sivan \& Y. Yaish for making their
data available to me. This work
  was supported by THE ISRAEL SCIENCE FOUNDATION  - Centers of Excellence
 Program,  and by the German Ministry of Science.

\end{multicols}
\end{document}